\documentclass[11pt,letterpaper]{article}

\usepackage[T1]{fontenc}
\usepackage[margin=1in]{geometry}
\usepackage{amsmath,amssymb,amsthm,mathtools}
\usepackage{wrapfig}
\usepackage{xurl}
\usepackage{hyperref}

\DeclareMathOperator{\E}{\mathbb{E}}
\renewcommand{\Pr}{\mathbb{P}}
\newcommand{\Oracle}{\mathcal{O}}
\newcommand{\TOK}{\operatorname{TOK}}
\newcommand{\noopsort}[1]{}

\newtheorem{theorem}{Theorem}
\newtheorem{lemma}{Lemma}
\newtheorem{proposition}{Proposition}

\theoremstyle{remark}

\title{Computing Weighted Set Covers with Heterogeneous Stochastic Oracles:
Deadline-Constrained Token Complexity}
\author{Jie Wang\thanks{Richard R. Miner School of Computer and Information
Sciences, the University of Massachusetts, Lowell, MA 01854, USA.}\\
\texttt{Jie\_Wang@uml.edu}}
\date{}

\begin{document}

\begin{titlepage}
\maketitle
\thispagestyle{empty}

\begin{abstract}
We study computation in which a probabilistic Turing machine (PTM) uses AI
systems, abstracted as heterogeneous stochastic oracles (HSOs), to solve Weighted
Set Cover instances, an approach supported by a pilot study showing that
an LLM can outperform weighted greedy. The PTM requests a cover for specified
elements and checks whether the returned sets cover those elements. HSOs
differ in capabilities, response distributions, latencies, and token costs.
We therefore study how the PTM selects and queries HSOs to obtain a full cover
satisfying a prescribed weight threshold by a prescribed completion deadline
and with a required success probability, and analyze the resulting expected
token cost.
For full-cover queries, with call successes independent and call failures
uninformative, we use an LP to characterize the minimum expected
token cost for each feasible success probability and show that an optimum
randomizes over at most two preselected sequences of oracle batches. Among
policies attaining the maximum success probability, we show that the expected token-cost reduction
of two-batch selection relative to one-batch all-oracle selection is a
nonnegative submodular function, and construct an FPTAS that returns a
two-batch policy with near-optimal cost reduction under heterogeneous
response-time bounds. We then show that computing an optimal
two-batch selection is NP-hard, but this optimization admits a pseudopolynomial
algorithm when expected token costs are positive integers. We next consider
partial-cover responses. The PTM combines the returned sets across batches and
turns and, once their union covers the full universe, verifies the assembled cover's
total weight directly or derives a weight guarantee from LP-dual tests applied
to the retained partial covers. We present one-turn probability,
token-cost, and hardness results, derive token-cost bounds for adaptive
selection under per-batch assumptions, and demonstrate with a constructed instance that residual-guided
querying can reduce expected token cost by a factor linear in the number of
possible residual subproblems.
\end{abstract}
\end{titlepage}
\setcounter{page}{1}

\section{Introduction}

Weighted Set Cover (WSC) is NP-hard~\cite{GareyJohnson1979}, and the
logarithmic worst-case guarantee of the weighted-greedy (WG)
algorithm~\cite{Chvatal1979} is essentially
optimal~\cite{DinurSteurer2014}; nevertheless, lighter covers may be
still be computed efficiently on particular instances. Our pilot study shows that an LLM can
outperform WG on a substantial fraction of the tested
instances, supporting the AI-augmented approach studied in this paper.

The query-response behavior of an LLM or any other query-responsive AI system,
such as a goal-directed agentic system, can be abstracted as a stochastic oracle. This
leads to an extension of the classical oracle
Turing-machine model: a probabilistic Turing machine (PTM) performs the local
computation and, together with a collection of heterogeneous stochastic
oracles (HSOs), forms a Stochastic-Oracle Turing Machine
(SOTM)~\cite{Wang2026comp,Wang2026agentic}. The HSOs may differ in their
capabilities, response distributions, latencies, and query-response token
costs. To assist its computation, the PTM chooses one or more HSOs, requests
a cover for specified elements, and independently verifies whether the returned
sets cover those elements. Oracle calls issued in one turn run concurrently. Across
turns, the PTM may use earlier responses to select later calls, stop after
accepting a cover, or retain returned sets and query HSOs for the uncovered
residual instance. We study how the PTM selects and queries HSOs to
obtain a full cover satisfying a prescribed weight threshold by a prescribed
completion deadline and with a required success probability, and analyze the
resulting expected token cost.

For full-cover queries, with call successes independent and call failures
uninformative, we use an LP to characterize the minimum expected token cost
for each feasible success probability and show that an optimum randomizes over
at most two preselected sequences of oracle batches. Among policies attaining
the maximum success probability, we show that the expected token-cost
reduction of two-batch selection relative to one-batch all-oracle selection is
a nonnegative submodular function, and construct an FPTAS that returns a
two-batch policy with near-optimal cost reduction under heterogeneous
response-time bounds. We then show that computing an optimal two-batch
selection is NP-hard, but this optimization admits a pseudopolynomial
algorithm when expected token costs are positive integers.

Our two-turn objective is closely related to the two-slot Time-Critical
Testing Problem, for which weak NP-hardness, a pseudopolynomial algorithm,
and an FPTAS for total expected cost are known
~\cite{AgnetisHermansLeusRostami2022}. That model uses unit-duration tests
and a fixed parallel capacity; ours permits heterogeneous call cutoffs and
variable batch sizes. Sequential and batch testing are
studied in~\cite{Unluyurt2004,DaldalGamzuSegevUnluyurt2016,
DaldalOzlukSelcukShahmoradiUnluyurt2017,SegevShaposhnik2022,TanXuNagarajan2025}.
Our full-cover results characterize the cost--success tradeoff and approximate
the submodular token-cost saving, rather than merely total cost.

We next consider partial-cover responses. The PTM combines the returned sets
across batches and turns and, once their union covers the full universe,
verifies the assembled cover's total weight directly or derives a weight
guarantee from LP-dual tests applied to the retained partial covers. For
one-turn partial-cover selection, we give a success-probability bound and a
greedy token-cost guarantee relative to a sufficient failure-risk
benchmark. When the per-element certified coverage probability lower bound and
desired success probability are fixed constants, the cost factor is
$O(\log^2 n)$; it improves
to $O(\log n)$ when the desired success probability is
$1-n^{-\Theta(1)}$, where $n$ is the universe size.

For adaptive selection,
history-conditional assumptions yield explicit token-cost bounds without
response independence; the lower bound additionally limits the coverage
revealed by every batch. We also show how adaptivity can save tokens: rather
than querying in advance for every residual subproblem that might arise, the
PTM uses an initial response to identify the actual residual and issues one
targeted follow-up query. In a constructed setting, this reduces expected
token cost by a factor proportional to the number of possible residual
subproblems.

Our partial-cover results combine classical primal-dual Set Cover charging
~\cite{RajagopalanVazirani1998} and greedy submodular-cover analysis
~\cite{Wolsey1982} with history-conditional coverage guarantees, token costs,
and deadlines. Stochastic and adaptive covering has also been studied in
~\cite{GolovinKrause2011,AgarwalAssadiKhanna2019,GhugeGuptaNagarajan2024}.
Our bounds use local LP-dual tests on retained responses without assuming
independent item realizations or a known adaptive-submodular prior. The
linear-gap construction adapts the ``all but one'' pattern from stochastic
covering~\cite{GoemansVondrak2006}.

Section~\ref{sec:empirical} reports the pilot study,
Section~\ref{sec:model} defines the model, and
Sections~\ref{sec:certified-set-cover}--\ref{sec:residual-cover} establish the
full- and partial-cover results. Section~\ref{sec:conclusion} concludes.

\section{LLM Capabilities for WSC: A Pilot Study}
\label{sec:empirical}

Write a WSC instance as a nonempty universe $U$ of $n$ elements and candidate
sets $S_1,\ldots,S_m\subseteq U$ whose union covers $U$, with positive weights
$w_1,\ldots,w_m$. Write $\mathrm{OPT}$ for the minimum weight of an integral
cover.
In the pilot study described below, we found that an LLM returned a full cover
lighter than a locally computed WG or reverse-deletion (RD) cover in a
substantial fraction of the tested instances. In our comparisons, RD starts
from the WG cover, processes its sets in nonincreasing weight order (breaking
ties by decreasing set index), and removes a set whenever its deletion
preserves coverage. We used OR-Library railway crew-scheduling instances, whose universe
consists of crew-scheduling tasks and whose candidate sets are feasible crew
duties with weights 1 or 2~\cite{Beasley1990ORLibrary,CapraraFischettiToth1999},
and Defects4J test-coverage matrices, whose universe consists of source-code
lines and whose candidate sets contain the lines executed by individual
tests~\cite{JustJalaliErnst2014,QuantumTesting2026}. Because Defects4J provides
no per-test costs, we assigned unit weight to every candidate set.

Railway instances can contain nearly $5{,}000$ universe elements and about
one million candidate sets~\cite{CapraraFischettiToth1999}, making computing
exact optima impractical. We therefore reduced each instance by using a
pseudorandom generator to grow a connected set of $n_r$ universe elements,
projecting the source candidate sets onto those elements, and capping the
resulting candidate family at $m_r$ sets.
When projection caused distinct source candidate sets to cover the same subset
of these elements, we kept, for each such group, a least-cost representative,
breaking ties by increasing source-set identifier, and removed any candidate set
dominated by a no-more-expensive superset; these two operations preserve the
optimum of the projected instance. WG breaks equal cost-per-new-element ratios
by lower weight and then increasing source-set identifier. The reduced family
first included its WG cover, then the lowest-cost-per-element alternatives
until $\lfloor m_r/2\rfloor$ sets had been selected (unless WG already used
more), breaking ties by larger cardinality and then identifier; the remaining
slots up to $m_r$ were filled pseudorandomly.
For the initial experiment, we set $(n_r,m_r)=(18,24)$, producing 12 reduced
instances. Eight were derived from OR-Library \texttt{rail516}, and four from
Defects4J.

We divided the 12 reduced instances into three groups of four. Group 1 used
two Defects4J and two railway projections generated with seeds
$20261019$ through $20261022$, without screening for a gap between the optimum
and either local baseline.
Group 2 contained four railway instances whose exact optima were lighter
than their WG covers. Group 3 contained two railway and two Defects4J instances
whose exact optima were lighter than their RD covers. An exact dynamic program
computed each optimum, and all instance selection and baseline computations
preceded the model calls.

\begin{wraptable}[3]{R}{0.54\textwidth}
\vspace{-0.6\baselineskip}
\centering
\small
\begin{tabular}{@{}lrrr@{}}
Run & Optimal & w.r.t.\ WG & w.r.t.\ RD\\
\hline
Opus 5.5 ($m$) & $33/36$ & $26/10/0$ & $10/25/1$\\
Opus 5.5 ($h$) & $36/36$ & $27/9/0$ & $12/24/0$
\end{tabular}
\vspace{-1.0\baselineskip}
\end{wraptable}
We conducted two runs using Opus 5.5~\cite{AnthropicOpus55}, one at reasoning
effort \texttt{medium} ($m$) and the other at \texttt{high} ($h$), making three repeated
calls per instance: 36 calls per run and 72 in total. All 72 returned families
were locally verified as valid covers. In either comparison column, an entry
$a/b/c$ gives the numbers of calls returning covers lighter than, equal in
weight to, or heavier than the named baseline.
Median wall-clock latency per call was 7.04 and 7.65 seconds, and the costs
of the 36-call runs were \$0.439 and \$0.448, respectively.

For a larger follow-up, we set $(n_r,m_r)=(26,100)$. From railway projections
generated in a predetermined order, we retained the first eight having at
least 90 distinct nondominated candidate sets. We did not use Defects4J because it did
not meet these size requirements.
The resulting railway instances had between
92 and 100 candidate sets.
Across three Opus 5.5 calls at reasoning effort \texttt{medium} per instance, all 24
responses were valid covers and 20 were optimal. On instances admitting a
strict improvement, all 9 calls beat WG and 3 of 6 beat RD. Median wall-clock
latency was 9.54 seconds, and the 24 calls had a
total cost of \$0.646.
The pilot used a per-call runtime timeout and recorded timed-out calls as
unsuccessful; it did not implement a shared deadline across multiple batches.
An earlier synthetic pilot used four unscreened and four greedy-gap instances.
All 24 calls returned optimal covers lighter than WG, and 15 also
improved on WG followed by RD.
All datasets and the complete experimental artifact are available on
GitHub.\footnote{\url{https://github.com/wanguml-beep/set-cover-stochastic-oracle-artifact}.}

\section{Multi-HSO SOTM Computation for WSC}
\label{sec:model}

Throughout the remainder of the paper, we fix a WSC instance and a finite
collection of HSOs unless stated otherwise. The standard LP relaxation
minimizes $\sum_{j=1}^m w_jx_j$ over $x_j\geq0$, subject to
$\sum_{S_j\ni e}x_j\geq1$ for every $e\in U$; let $L$ denote its optimum.
Here $x_j$ represents the fractional extent to which $S_j$ is selected.
Because every integral cover is feasible for this relaxation,
$L\leq\mathrm{OPT}$.

Every oracle call for a full cover uses a fixed query $q$ asking an HSO to
return an index set $\mathcal C\subseteq[m]$ representing the subfamily
$\{S_j:j\in\mathcal C\}$.
Whether a proposed family covers $U$ can be checked in polynomial time.
The PTM verifies every returned family independently. For every cover
passing this check, it reports the \emph{instance-specific certified factor}
$\rho(\mathcal C)=w(\mathcal C)/L$, where
$w(\mathcal C)=\sum_{j\in\mathcal C}w_j$. Because $L\leq\mathrm{OPT}$,
\(w(\mathcal C)/\mathrm{OPT}\leq\rho(\mathcal C)\). Given a target factor
$\rho\geq1$, the PTM accepts $\mathcal C$ when
$w(\mathcal C)\leq\rho L$, equivalently $\rho(\mathcal C)\leq\rho$; we call such an accepted full cover a
\emph{certified $\rho$-cover}.
Sections~\ref{sec:one-turn-partial-cover} and~\ref{sec:residual-cover} also
allow queries for specified subsets of $U$. There the PTM may retain partial
proposals and accept their union once it forms a full cover satisfying
$w(\mathcal C)\leq\rho L$, either by checking this inequality directly or by
using the local LP-dual tests defined there. Thus, \emph{certified
$\rho$-cover} has the same meaning for full- and partial-cover responses.

Different configurations of the same underlying AI model, such as different
reasoning-effort settings, are treated as distinct HSOs because they may
differ in response distribution, response quality, token cost, or latency.
Let $M$ be a PTM with access to these HSOs. In an oracle call, the PTM sends a query
string to an oracle and receives a visible response string, which may contain
intermediate messages and a final message. The oracle's response distribution
may depend on the query, its internal state, its accessible environment, and
the prior transcript. The PTM can sample this distribution only by querying
the oracle.

In an \emph{oracle turn}, the PTM selects a nonempty batch of oracle calls and
forms every query before observing any response from the batch. The calls run
concurrently and continue until they return or are canceled, even if an
acceptable response arrives earlier; only then may the PTM select another
batch.

An oracle call has a latency, measured from its issuance until its response or
cancellation. The PTM may impose a \emph{cutoff}, canceling an unfinished call at a
predetermined elapsed time and treating the oracle--cutoff pair as one
selectable configuration. Batch latency is the maximum call latency.

Under a fixed finite retry limit, we associate each oracle with finitely many
labeled call opportunities, index all such opportunities by $[k]$, and allow
each to be used at most once. Multiple opportunities associated with the same
underlying oracle may be issued concurrently, for example through separate
sessions or API credentials. In
turn $t$, let $A_t\subseteq[k]$ be the random batch of selected opportunities.
Let $T$ be the random number of oracle turns executed before the PTM halts.
Section~\ref{sec:one-turn-partial-cover} uses an equivalent grouped notation
for homogeneous repeated calls: there $[k]$ indexes the distinct oracles, and
$\boldsymbol{\zeta}\in\mathbb Z_{\geq0}^k$ records the number of concurrent
calls issued to each oracle.

Fix a response deadline $D>0$ and a positive integer $u$ bounding the number
of oracle turns. Local computation time is excluded from $D$. A policy is
\emph{deadline- and turn-feasible} if it issues at most $u$ turns before $D$.
After each completed batch, the PTM either continues or halts. It halts when
that batch yields or completes an accepted full cover, its policy ends, $u$
turns have been issued, or the deadline is reached. At $D$, every unfinished
call is canceled and treated as unsuccessful; if no accepted full cover is
available, the PTM reports failure.

A deadline- and turn-feasible policy is \emph{nontruncating} if every issued
batch can run through its prescribed cutoff or almost-sure latency upper bound
before $D$. For a batch $A$ with individual cutoffs $\ell_i$, write
$\ell_A=\max_{i\in A}\ell_i$; nontruncation requires the sum of the allotted
batch durations along every possible batch sequence to be at most $D$.
Sections~\ref{sec:one-turn-partial-cover} and~\ref{sec:residual-cover} use
their stated almost-sure batch-latency upper bounds as the allotments. The
cutoff policies in Section~\ref{sec:certified-set-cover} and the constructive
upper-bound policies below are nontruncating. The lower bound in
Theorem~\ref{thm:bounded-patch-sandwich}\textup{(2)} applies to the broader
class of deadline- and turn-feasible policies and hence also to nontruncating
ones.

Each oracle tokenizes the query and visible response internally, possibly
using a different tokenizer. An oracle call has a nonnegative random
\emph{token cost}, equal to the query, visible-response, and billed
hidden-reasoning tokens charged for the call. The vendor reports this cost. Batch token
costs add, and a canceled call incurs the cost accrued through cancellation.
For every reached turn $t$ and $i\in A_t$,
let $\TOK_{t,i}$ be the random token cost of opportunity $i$. The total token
cost is $\TOK_M=\sum_{t=1}^{T}\sum_{i\in A_t}\TOK_{t,i}$; local computation is not
charged.
Given an acceptance criterion and a class of oracle-selection policies,
fix a required probability $\eta\in(0,1]$ that the oracle-selection phase
produces an accepted cover by $D$. The \emph{instance-level token complexity} of the resulting selection
problem is the infimum of $\E[\TOK_M]$ over PTMs following policies in that
class that are deadline- and turn-feasible and satisfy the probability
requirement.
It is $+\infty$ when no such
PTM exists. The expectation is over the PTM's internal random source and the
joint randomness of the oracle calls.

For each full-cover call opportunity $i\in[k]$ in Section~\ref{sec:certified-set-cover},
let $\ell_i>0$ be its cutoff,
$p_i\in(0,1)$ the probability that it returns an accepted cover by $\ell_i$,
and $c_i>0$ its expected token cost through response or cancellation at
$\ell_i$. The realized token cost may correlate with success. The PTM chooses
the oracle configuration, query, and cutoff, but does not control the realized
response quality, latency, or billed hidden-reasoning cost. In applications,
$p_i$ and $c_i$ may be estimated from prior calls and provider usage reports
for similar instances. The strict upper bound on $p_i$ ensures that every
all-failure schedule used below has positive probability; certain-success
calls require a separate characterization.

Whenever numerical parameters are supplied as algorithmic input, they are
assumed to be rational and encoded in binary; parameters in purely analytic
statements may be arbitrary real values in their stated ranges.

We call a PTM's oracle-selection policy \emph{selection-only} if every cover
counted as a success consists solely of sets proposed by queried oracles,
either as one full proposal or as a union of partial proposals. The PTM may still
compute locally to formulate queries, select oracles, solve the LP relaxation,
and verify proposed covers and approximation certificates.
After this phase terminates, the PTM may return the lightest among any
HSO-derived cover and a locally computed WG cover improved by RD. This
post-hoc comparison does not change the selection-only success event or its
token complexity.

\section{Oracle Selection with Full-Cover Responses}
\label{sec:certified-set-cover}

\phantomsection
\label{ass:fica}
Earlier failed batches do not change the behavior of calls not yet
made. Specifically, after any sequence of failed batches, each unused call
opportunity $i$ still has success probability $p_i$ and expected token cost
$c_i$, and the success events of subsequently issued calls are conditionally
independent. We call this the \emph{failure-invariant call assumption} (FICA).

After an issued batch fails, a \emph{failure-oblivious} selection-only policy
chooses the next batch using only the identities of the call opportunities
already used and the PTM's random source. It ignores the returned text,
realized token costs, and completion times of unsuccessful calls. Throughout
this section, we assume the FICA and restrict attention to failure-oblivious
selection-only policies that are nontruncating. Each batch is allotted the
maximum cutoff among its calls, regardless of whether all calls finish earlier.

For a batch $A\subseteq[k]$, write $c_A=\sum_{i\in A}c_i$,
$f_A=\prod_{i\in A}(1-p_i)$, and $\ell_A=\max_{i\in A}\ell_i$ for its
expected token cost, all-failure probability, and allotted duration,
respectively, with $c_\varnothing=\ell_\varnothing=0$ and
$f_\varnothing=1$. If the calls indexed by $A$ are made first and
those indexed by $[k]\setminus A$ are made only after $A$ fails, the
expected total token cost is $c_A+f_A\cdot c_{[k]\setminus A}$.
Let $p^*=1-\prod_{i=1}^k(1-p_i)$ be the probability that at least one call
succeeds when all $k$ call opportunities are used. These quantities give the
exact one- and two-turn optima below.

\begin{proposition}
\label{prop:certified-set-cover}
Suppose $\ell_{[k]}\leq D$. Among policies under which the PTM attains the
maximum success probability $p^*$, the minimum expected token cost with at
most one oracle turn is $c_{[k]}$. With at most two oracle turns, it is the minimum of
$c_A+f_A\cdot c_{[k]\setminus A}$ over nonempty $A\subseteq[k]$ satisfying
$\ell_A+\ell_{[k]\setminus A}\leq D$, with the empty second batch omitted
when $A=[k]$.
\textup{(See Appendix~\ref{app:proof-certified-set-cover} for a proof.)}
\end{proposition}

We now consider arbitrary turn bounds and required success probabilities. The
\emph{all-failure schedule} of a selection-only policy is the ordered sequence
of batches that the PTM would follow if no call ever returned an accepted
cover. For a failure-oblivious policy, this schedule depends only on the PTM's random source;
denote the resulting random schedule by $\mathbf B$. Let $\mathcal B$ be the
finite set of sequences $B=(B_1,\ldots,B_t)$, where $0\leq t\leq u$, the
$B_j$ are pairwise disjoint nonempty subsets of $[k]$, and
$\sum_{j=1}^t\ell_{B_j}\leq D$. For $t=0$, $B$ is the empty schedule.
For $B\in\mathcal B$, let
$P_B=1-\prod_{j=1}^t f_{B_j}$ be the probability that the PTM returns an
accepted cover while following $B$, and let
$c(B)=\sum_{j=1}^t c_{B_j}\prod_{s<j}f_{B_s}$ be the expected token cost
incurred while following $B$. For the empty schedule,
$P_B=c(B)=0$. Two selection-only policies are \emph{equivalent} if they
have the same success probability and expected token cost. Fix a required
success probability $\eta\in(0,1]$.

\begin{theorem}
\label{thm:general-cost-frontier}
\textup{(1)} Every randomized \emph{failure-oblivious} nontruncating
selection-only policy $\pi$ with deadline $D$ and turn bound $u$ has an equivalent
selection-only policy $\widehat{\pi}$ that samples a schedule
$B\in\mathcal B$ before its first oracle call and follows $B$ until a batch
returns an accepted cover or the schedule ends.

\textup{(2)} Under the rational-input convention, the infimum defining
instance-level token complexity is attained and equals
$\min\{\sum_{B\in\mathcal B}w_B\cdot c(B):
\sum_{B\in\mathcal B}w_B\cdot P_B\geq\eta,\
\sum_{B\in\mathcal B}w_B=1,\
w_B\geq0\text{ for }B\in\mathcal B\}$, with the minimum of an empty feasible
set interpreted as $+\infty$.

\textup{(3)} If $\eta\leq\max_{B\in\mathcal B}P_B$, an optimal policy
can be implemented by having the PTM select, before its first oracle call,
one of at most two schedules according to a probability distribution.
\end{theorem}
\begin{proof}
\textup{(1)} Conditioning on a realization of $\pi$'s random source, which is
independent of the oracle responses, makes the PTM's oracle-selection
decisions deterministic. Because the PTM continues
only after failure and later selections depend on no other response
information, its all-failure execution follows some
$B\in\mathcal B$. It reaches batch $B_j$ exactly when all preceding
batches fail, an event of probability $\prod_{s<j}f_{B_s}$. Independence of
the call-success events therefore gives success probability $P_B$.
The FICA and linearity of expectation give expected
token cost $c(B)$. Writing $w_B=\Pr_\pi[\mathbf B=B]$, taking the weighted
average over $B\in\mathcal B$ gives success probability
$\sum_Bw_B\cdot P_B$ and expected token cost $\sum_Bw_B\cdot c(B)$. Before
making a call, the PTM can run its randomized selection procedure on a
hypothetical history in which every batch fails, thereby determining the
entire schedule with this induced distribution. Following the resulting
schedule defines $\widehat{\pi}$ and proves \textup{(1)}.

\textup{(2)} The preceding argument maps every feasible policy to a
feasible distribution with the same objective value. Under the rational-input
convention, the linear program in statement~\textup{(2)} has rational data,
and its feasible region is a compact polytope. Whenever it is nonempty, an optimum is attained
at a rational vertex. Each coordinate of this vertex is indexed by a schedule
$B\in\mathcal B$; because the coordinates are nonnegative and sum to $1$,
the vertex specifies a rational probability distribution over schedules.
Before its first call, the PTM can sample this distribution exactly from its
random source and then follow the sampled schedule. It halts after the first batch that
returns an accepted cover or after the schedule ends, attaining success
probability $\sum_Bw_B\cdot P_B$ and expected token cost
$\sum_Bw_B\cdot c(B)$ within the deadline and turn bound. If the feasible set
is empty, \textup{(1)} shows that no policy meets the success requirement, so
both sides are $+\infty$. This proves \textup{(2)}.

\textup{(3)} Choose an optimal vertex $w^*=(w_B^*)_{B\in\mathcal B}$ of
the linear program in statement~\textup{(2)}. Suppose that at least three
coordinates of $w^*$ are positive. Restricted to these coordinates, the two
homogeneous equations $\sum_B\delta_B=0$ and
$\sum_B P_B\delta_B=0$ have a nonzero solution
$\boldsymbol{\delta}=(\delta_B)_{B\in\mathcal B}$ such that $\delta_B=0$
whenever $w_B^*=0$. For sufficiently small $\varepsilon>0$, both
$w^*+\varepsilon\boldsymbol{\delta}$ and
$w^*-\varepsilon\boldsymbol{\delta}$ are nonnegative,
$\sum_B(w_B^*\pm\varepsilon\delta_B)=\sum_Bw_B^*=1$, and
$\sum_BP_B(w_B^*\pm\varepsilon\delta_B)=\sum_BP_Bw_B^*\geq\eta$.
Thus, $w^*+\varepsilon\boldsymbol{\delta}$ and
$w^*-\varepsilon\boldsymbol{\delta}$ are distinct feasible points, while
$w^*$ is their midpoint. A vertex cannot be a nontrivial convex combination
of two distinct feasible points, so this contradicts that $w^*$ is a vertex.
Hence, at most two coordinates of $w^*$ are positive. Since each positive
coordinate corresponds to a schedule selected with positive probability, the
PTM selects among at most two schedules.
\end{proof}

Theorem~\ref{thm:general-cost-frontier} is a structural characterization:
$|\mathcal B|$ may be exponential in $k$ and $u$. We next specialize to two
turns at maximum attainable success to obtain an FPTAS.

Suppose $\ell_{[k]}\leq D$. Let $c_2^*$ be the minimum expected token cost,
among selection-only policies, of attaining success probability $p^*$ by $D$
with at most two oracle turns, and define the saving function
$F(A)=(1-f_A)c_{[k]\setminus A}$.
For nonempty $A\subseteq[k]$ satisfying
$\ell_A+\ell_{[k]\setminus A}\leq D$, let $\pi_A$ query $A$ first and, if
that batch fails, query $[k]\setminus A$, omitting any empty batch. Call such
an $A$ \emph{two-batch feasible}.

\begin{theorem}
\label{thm:set-cover-savings}
\textup{(1)} The function $F$ is nonnegative and submodular.

\textup{(2)} There is an FPTAS for maximizing $F(A)$ over nonempty
two-batch-feasible choices. For every $\varepsilon\in(0,1)$, it constructs a policy
$\pi_A$ in deterministic time polynomial in the total encoding length of $D$
and the $k$ triples $(p_i,c_i,\ell_i)$, and in $1/\varepsilon$. This policy attains
success probability $p^*$ by $D$
and satisfies
$F(A)\geq(1-\varepsilon)(c_{[k]}-c_2^*)$ and
$c_A+f_A\cdot c_{[k]\setminus A}\leq c_2^*+\varepsilon(c_{[k]}-c_2^*)$.
\end{theorem}
\begin{proof}
\textup{(1)} The savings identity follows from
$c_A+f_A\cdot c_{[k]\setminus A}=c_{[k]}-F(A)$. The set function $F$ is
nonnegative because $0\leq f_A\leq1$ and $c_{[k]\setminus A}\geq0$. By the
diminishing-returns characterization of submodularity, it remains to compare
marginal gains. Take $A\subseteq A'\subseteq[k]$ and $i\notin A'$, and put
$q_X=f_X$ and $h_X=c_{[k]\setminus X}$ for $X\in\{A,A'\}$. Then
\(F(A\cup\{i\})-F(A)
=p_iq_A(h_A-c_i)-c_i(1-q_A)\).
Because $i\notin A'$, $h_A-c_i\geq h_{A'}-c_i\geq0$. Together with
$q_A\geq q_{A'}$ and $1-q_A\leq1-q_{A'}$, this shows that the marginal
gain is at least the corresponding marginal gain at $A'$. Hence $F$ is
submodular.

\textup{(2)} The approximation scheme uses the product--sum form of $F$.
Enumerate $a\in\{\ell_1,\ldots,\ell_k\}$ and
$b\in\{0,\ell_1,\ldots,\ell_k\}$ with $a+b\leq D$. For each pair, let
$L_{a,b}=\{i:\ell_i>b\}$, $U_a=\{i:\ell_i\leq a\}$, and
$\mathcal A_{a,b}=\{A\subseteq[k]:L_{a,b}\subseteq A\subseteq U_a\}$.
Whenever $L_{a,b}\subseteq U_a$, every nonempty member of
$\mathcal A_{a,b}$ is two-batch feasible. Conversely, every
two-batch-feasible $A$ belongs to one such family
by taking
$a=\ell_A$ and $b=\ell_{[k]\setminus A}$: every oracle whose cutoff
exceeds $b$ must then belong to $A$, so
$L_{a,b}\subseteq A\subseteq U_a$.

Fix one such family and abbreviate $L=L_{a,b}$, $U=U_a$,
$E=[k]\setminus U$, and $S=U\setminus L$. Choosing the portion
$B\subseteq S$ placed in the second batch determines
$A=L\cup(S\setminus B)$; the complete second batch is $E\cup B$. Put
$q_i=1-p_i$, $\bar q_B=\prod_{i\in S\setminus B}q_i$, and
$c_B=\sum_{i\in B}c_i$. Since $f_A=f_L\bar q_B$ and
$[k]\setminus A=E\cup B$, $F(A)=(c_E+c_B)(1-f_L\bar q_B)$; denote this
expression by $G(B)$. For a fixed value of $c_B$, $G(B)$ is nonincreasing in
$\bar q_B$. We therefore round $c_B$ and, for each rounded value, retain a
subset minimizing $\bar q_B$.

We construct an approximation for this family. Include $B=\varnothing$
as an explicit candidate.
For each $r\in S$, consider solutions in which $r$ is a maximum-cost member
of $B$. Restrict the eligible items to
$S_r=\{i\in S:c_i\leq c_r\}$, require $r\in B$, and set
$K=\varepsilon c_r/k$ and $z_i=\lfloor c_i/K\rfloor$. Initialize a dynamic
program with $r$ selected, rounded cost $z_r$, and factor
$\prod_{i\in S\setminus S_r}q_i$. When processing
$i\in S_r\setminus\{r\}$, selecting $i$ adds $z_i$ to the rounded cost,
whereas omitting it multiplies the product by $q_i$; for each attainable
rounded cost, retain the smaller product and its subset. The resulting
product is $\bar q_B$. There are at most $k^2/\varepsilon$ attainable sums.
Rational products can be compared exactly with bit length polynomial in the
input. Evaluate $G(B)$ for every stored subset whose corresponding first
batch $A$ is nonempty.

Let $B^*\subseteq S$ maximize $G(B)$, and set
$A^*=L\cup(S\setminus B^*)$.
If $B^*=\varnothing$, this candidate was evaluated directly. Otherwise, choose a
maximum-cost $r\in B^*$. The corresponding dynamic program considers
$B^*$. Let $B'$ be the stored subset with the same rounded sum and minimum
product, and set $A'=L\cup(S\setminus B')$. Then
$\bar q_{B'}\leq\bar q_{B^*}$. If $L=\varnothing$, the nonempty first batch
$A^*$ gives $S\setminus B^*\ne\varnothing$ and, since $p_i>0$,
$\bar q_{B^*}<1=\bar q_S$; hence $B'\ne S$. Thus $A'$ is nonempty in all
cases. Moreover,
$c_{B'}\geq K\sum_{i\in B'}z_i=K\sum_{i\in B^*}z_i
\geq c_{B^*}-kK\geq(1-\varepsilon)c_{B^*}$, where the last inequality
uses $kK=\varepsilon c_r\leq\varepsilon c_{B^*}$ because $r\in B^*$.
Thus $c_E+c_{B'}\geq(1-\varepsilon)(c_E+c_{B^*})$, while the second factor
of $G(B')$ is at least that of $G(B^*)$. Hence
$F(A')=G(B')\geq(1-\varepsilon)F(A^*)$ for
this family. Enumerating the at most $(k+1)^2$ families and returning
the best candidate is polynomial in the input length and $1/\varepsilon$
and achieves the same guarantee over all two-batch-feasible choices.
Proposition~\ref{prop:certified-set-cover} identifies the optimum saving as
$c_{[k]}-c_2^*$, proving the stated guarantees.
The selected schedule queries every oracle on its all-failure path and is
nontruncating, so it attains success probability $p^*$. Thus the
constructed policy satisfies all claims in \textup{(2)}.
\end{proof}

The FPTAS approximates the saving $c_{[k]}-c_2^*$ multiplicatively, which is
stronger than only approximating total expected cost when $c_2^*$ is close to
$c_{[k]}$.

Differences among oracle success probabilities, token costs, and cutoffs can
make an additional oracle turn especially valuable. For example, consider two
oracles with $(p_1,c_1,\ell_1)=(0.8,1,2)$ and
$(p_2,c_2,\ell_2)=(0.5,5,1)$. Their maximum success probability is $0.9$.
At deadline $D=2$, attaining success probability $0.9$ requires querying
both oracles in one batch, at expected cost $6$. At $D=3$, the optimal
two-turn policy queries $\Oracle_1$ first and $\Oracle_2$ only after failure,
attaining the same success probability at expected cost $1+0.2(5)=2$. This
example motivates studying the computational complexity of finding an optimal
two-turn policy at maximum attainable success.

The \emph{two-turn maximum-success selection problem}, denoted by
\textsc{2T-MaxSuccess}, has as input $k$ call opportunities with parameters
$(p_i,c_i,\ell_i)$ and a deadline $D$ satisfying $\ell_{[k]}\leq D$. Its
output is a two-batch-feasible choice $A\subseteq[k]$ minimizing
$c_A+f_A\cdot c_{[k]\setminus A}$, where the resulting policy queries $A$
first and, upon failure, queries $[k]\setminus A$, omitting any empty batch.
This policy attains the
maximum success probability $p^*$.

With uniform cutoffs, this objective resembles two-slot time-critical testing
~\cite{AgnetisHermansLeusRostami2022}. The next theorem adapts its hardness
and dynamic-programming ideas to our setting and permits heterogeneous cutoffs.

\begin{theorem}
\label{thm:two-turn-exact-hardness}
\textup{(1)} \textsc{2T-MaxSuccess} is NP-hard even when all cutoffs and all
success-per-cost ratios are equal.

\textup{(2)} If the $c_i$ are positive integers, \textsc{2T-MaxSuccess} can
be solved in time polynomial in the input length and $c_{[k]}$, with
success probabilities and cutoffs that may differ among call opportunities.
\textup{(See Appendix~\ref{app:proof-two-turn-hardness} for a proof.)}
\end{theorem}

The reduction from \textsc{Partition} uses uniform cutoffs and
success-per-cost ratios, showing that hardness arises from the nonlinear
conditional second-batch cost $f_A\cdot c_{[k]\setminus A}$. The
pseudopolynomial algorithm enumerates cutoff pairs and applies subset-sum
dynamic programming. Small $k$ alone need not make it practical, because
billed hidden-reasoning tokens may make $c_{[k]}$ large; direct enumeration of
the $2^k$ first batches avoids this dependence when $k$ is sufficiently small.

\section{One-Turn Selection with Partial-Cover Responses}
\label{sec:one-turn-partial-cover}

Partial-cover responses may arise from incomplete solutions or from queries
deliberately restricted to subinstances. In one oracle turn, the PTM may
assemble a cover from several such responses. Let $\mathcal C\subseteq[m]$ be the indices of the sets retained by
the PTM, and let $R=U\setminus\bigcup_{j\in\mathcal C}S_j$ be the residual set
of uncovered elements. The dual of the LP relaxation has the same optimum $L$:
\(L=\max\{\sum_{e\in U}y_e:y_e\geq0,
\sum_{e\in S_j}y_e\leq w_j\text{ for every }j\in[m]\}\).
Fix a feasible dual vector $\mathbf y=(y_e)_{e\in U}$. For a returned
subfamily $P\subseteq[m]$, let
$N(P,R)=R\cap\bigcup_{j\in P}S_j$ be the elements of $R$ newly covered by $P$. We say that
$P$ satisfies the \emph{local dual-certification condition} for $R$ if
$N(P,R)\ne\varnothing$ and
\(w(P):=\sum_{j\in P}w_j\leq\rho\sum_{e\in N(P,R)}y_e\).
The PTM retains $P$ only if this condition holds, applying primal-dual
charging to oracle responses~\cite{RajagopalanVazirani1998}.

\begin{lemma}
\label{lem:patch-composition}
Suppose the PTM successively retains partial covers $P_1,\ldots,P_H$ and,
for each $h\in[H]$, $P_h$ satisfies the local dual-certification condition for
$R_{h-1}=U\setminus\bigcup_{\ell<h}\bigcup_{j\in P_\ell}S_j$. If the sets
indexed by $\bigcup_{h=1}^H P_h$ cover $U$, the resulting cover
$\mathcal C=\bigcup_{h=1}^H P_h$ is a certified $\rho$-cover satisfying
$w(\mathcal C)\leq\rho L\leq\rho\,\mathrm{OPT}$.
\textup{(See Appendix~\ref{app:proof-patch-composition} for a proof.)}
\end{lemma}

Assume that the $a$th call to oracle $i$ returns a random partial cover. For
each oracle $i$, let $V_i\subseteq U$ be a known set of elements for which its
calls have the following guarantee. Fix an order in which the PTM processes
the responses after all calls finish. For every $e\in V_i$, every call to
oracle $i$, and every positive-probability history of earlier processed
responses with $e\in R$, assume that the call's partial cover both satisfies
the local dual-certification condition for $R$ and covers $e$ with conditional
probability at least $p>0$. The realized token cost $\TOK_{i,a}>0$ may be
random, and the known bound
$\TOK_{i,a}\leq c_i$ is assumed to hold almost surely. Each call has
latency at most $D$.

This assumption is processing-order sensitive in general. One sufficient
order-robust condition is that calls are independent and, for every $e\in V_i$,
a response from oracle $i$ contains with probability at least $p$ a subfamily
$P_e$ covering $e$ with $w(P_e)\leq\rho y_e$. If $e$ is still uncovered when
the response is processed, then $e\in N(P_e,R)$, so $P_e$ satisfies the local
dual-certification condition regardless of earlier coverage overlap.

For a vector $\boldsymbol{\zeta}\in\mathbb Z_{\geq0}^k$, the PTM issues $\zeta_i$ parallel calls
to oracle $i$. Write
$d_e(\boldsymbol{\zeta})=\sum_{i:e\in V_i}\zeta_i$ and
$c(\boldsymbol{\zeta})=\sum_i c_i\zeta_i$, the resulting upper bound on the batch's token cost.
We call $\boldsymbol{\zeta}$ an $r$-fold multicover if
$d_e(\boldsymbol{\zeta})\geq r$ for every $e\in U$; the case $r=1$ is an ordinary cover.
Assume $\bigcup_iV_i=U$ and $p,\eta\in(0,1)$. Set $q=1-p$, and let
$\mathrm{OPT}^{\mathrm{risk}}_\eta$ be the minimum $c(\boldsymbol{\zeta})$ over
vectors satisfying $\sum_{e\in U}q^{d_e(\boldsymbol{\zeta})}\leq1-\eta$. Note that
$\sum_{e\in U}q^{d_e(\boldsymbol{\zeta})}$ bounds the probability that some element
remains uncovered; thus, $\mathrm{OPT}^{\mathrm{risk}}_\eta$ is the minimum
batch-cost bound that reduces this certified failure risk to at most $1-\eta$. Set
$r_0=\lceil\ln(1/(1-\eta))/\ln(1/q)\rceil$, and
$r_1=\lceil\ln(n/(1-\eta))/\ln(1/q)\rceil$.
Then $1\leq r_0\leq r_1$.

\begin{theorem}
\label{thm:patch-policy-construction}
For every call-count vector $\boldsymbol{\zeta}$, the PTM returns a certified
$\rho$-cover by $D$ with probability at least
$\max\{0,1-\sum_{e\in U}q^{d_e(\boldsymbol{\zeta})}\}$ and incurs token cost at most
$c(\boldsymbol{\zeta})$ almost surely. Moreover, a greedy algorithm computes, in
time polynomial in $n$, $k$, and the numerical value of $r_1$, a vector
$\boldsymbol{\zeta}'$ whose one-turn batch returns a certified $\rho$-cover by
$D$ with probability at least $\eta$ and has token cost at most
$(1+\ln(nr_1))\lceil r_1/r_0\rceil\mathrm{OPT}^{\mathrm{risk}}_\eta$.
\end{theorem}
\begin{proof}
Intuitively, each call to an oracle $i$ gives every element in $V_i$ another
chance to be covered by a certified partial cover. Any selection satisfying
the failure-risk bound must give every element at least $r_0$ such chances,
whereas $r_1$ chances per element suffice to cover all $n$ elements with
probability at least $\eta$. Thus, the PTM constructs a one-turn batch with
the required success probability by finding a low-cost $r_1$-fold multicover
of the sets $V_i$, while an optimal $r_0$-fold multicover provides a lower
bound for comparison.

We first prove the guarantee for an arbitrary call-count vector.
Fix an element $e$. The one-turn batch contains $\zeta_i$ calls to each oracle
$i$, all issued before any response is observed; among them,
$d_e(\boldsymbol{\zeta})$ calls are to oracles $i$ with $e\in V_i$. Process the
completed responses in the fixed order. Conditional on $e$ remaining uncovered
before the response to any such call is processed, that response passes the local
dual-certification condition and covers $e$ with probability at least $p$.
Put $d=d_e(\boldsymbol{\zeta})$. For $j\in[d]$, let $E_j$ be the event that
$e$ is uncovered immediately before the $j$th such response is processed,
and let $A_j$ be the event that it remains uncovered immediately afterward.
If $d=0$, then $q^d=1$, so
$\Pr[e\text{ remains uncovered}]\leq q^d$ holds trivially. Suppose $d\geq1$.
Averaging the history-conditional guarantee gives
$\Pr[A_j\mid E_j]\leq q$ whenever $\Pr[E_j]>0$. Since coverage is retained,
$A_j\subseteq E_j$, so $\Pr[A_j]\leq q\Pr[E_j]$ also when
$\Pr[E_j]=0$. Moreover, $\Pr[E_1]\leq1$ and
$E_j\subseteq A_{j-1}$ for $j\geq2$; an intervening response may cover $e$.
Thus $\Pr[A_1]\leq q$ and
$\Pr[A_j]\leq q\Pr[A_{j-1}]$, yielding $\Pr[A_d]\leq q^d$.
After all responses are processed, the probability that $e$ remains
uncovered is at most $\Pr[A_d]$. By the union bound, the
retained partial covers cover $U$ with probability at least
$\max\{0,1-\sum_{e\in U}q^{d_e(\boldsymbol{\zeta})}\}$. On this event,
Lemma~\ref{lem:patch-composition} shows that their union is a certified
$\rho$-cover. All calls finish by $D$, and their almost-sure cost bounds sum to
$c(\boldsymbol{\zeta})$.

We next construct the vector $\boldsymbol{\zeta}'$.
Every vector feasible for the definition of
$\mathrm{OPT}^{\mathrm{risk}}_\eta$ satisfies
$q^{d_e(\boldsymbol{\zeta})}\leq1-\eta$ for each $e$, so
$d_e(\boldsymbol{\zeta})\geq r_0$. Conversely, every $r_1$-fold multicover satisfies
$\sum_e q^{d_e(\boldsymbol{\zeta})}\leq nq^{r_1}\leq1-\eta$. It therefore suffices
to construct a low-cost $r_1$-fold multicover.
Following the greedy submodular-cover algorithm~\cite{Wolsey1982}, view each
$V_i$ as a set that may be selected repeatedly at cost $c_i$ per selection,
and require every element to belong to the selected multiset with multiplicity
at least $r_1$. Let
$Q=nr_1$, $\Phi(\boldsymbol{\zeta})=\sum_{e\in U}\min\{r_1,d_e(\boldsymbol{\zeta})\}$, and
$\Delta_i(\boldsymbol{\zeta})=\Phi(\boldsymbol{\zeta}+\mathbf e_i)-\Phi(\boldsymbol{\zeta})$.
Then $\Delta_i(\boldsymbol{\zeta})
=|\{e\in V_i:d_e(\boldsymbol{\zeta})<r_1\}|$. Hence, if
$\boldsymbol{\zeta}\leq\boldsymbol{\xi}$ coordinatewise,
$\Delta_i(\boldsymbol{\zeta})\geq\Delta_i(\boldsymbol{\xi})$, so $\Phi$ has
diminishing returns.
Starting with $\boldsymbol{\zeta}=\mathbf 0$, the greedy algorithm chooses,
while $\Phi(\boldsymbol{\zeta})<Q$, an oracle $i$ with
$\Delta_i(\boldsymbol{\zeta})>0$ minimizing
$c_i/\Delta_i(\boldsymbol{\zeta})$, and increases $\zeta_i$ by $1$.
Let $\boldsymbol{\zeta}'$ be the vector at termination. Because
$\bigcup_iV_i=U$, an oracle with positive marginal gain exists whenever
$\Phi(\boldsymbol{\zeta})<Q$. Each iteration increases $\Phi$ by at least $1$,
so the algorithm terminates after at most $Q$ iterations. At termination,
$\Phi(\boldsymbol{\zeta}')=Q=nr_1$; since each of its $n$ summands is at most
$r_1$, every summand equals $r_1$, and $\boldsymbol{\zeta}'$ is an $r_1$-fold
multicover. Each marginal can be computed by scanning $V_i$, which proves the
stated runtime.

Let $\mathrm{OPT}_r$ denote the minimum cost of an $r$-fold multicover. At an intermediate vector
$\boldsymbol{\zeta}$, put $z=Q-\Phi(\boldsymbol{\zeta})$ and let $\boldsymbol{\zeta}^*$ be an optimal
$r_1$-fold multicover. Since $\Phi(\boldsymbol{\zeta}+\boldsymbol{\zeta}^*)=Q$,
adding the coordinates of $\boldsymbol{\zeta}^*$ one at a time and using
diminishing returns gives
$z=\Phi(\boldsymbol{\zeta}+\boldsymbol{\zeta}^*)-\Phi(\boldsymbol{\zeta})
\leq\sum_i \zeta_i^*\Delta_i(\boldsymbol{\zeta})$. Since
$\sum_i c_i\zeta_i^*=\mathrm{OPT}_{r_1}$, taking the minimum of
$c_i/\Delta_i(\boldsymbol{\zeta})$ over the indices with
$\zeta_i^*\Delta_i(\boldsymbol{\zeta})>0$ shows that some oracle satisfies
$c_i/\Delta_i(\boldsymbol{\zeta})\leq\mathrm{OPT}_{r_1}/z$, and greedy chooses a ratio no
larger. If a greedy step reduces the number of unmet requirements from $z$
to $z-\Delta$, its cost is at most
$\Delta\mathrm{OPT}_{r_1}/z\leq
\mathrm{OPT}_{r_1}\sum_{s=z-\Delta+1}^z1/s$. Summing over the steps gives
$c(\boldsymbol{\zeta}')\leq H_Q\mathrm{OPT}_{r_1}
\leq(1+\ln(nr_1))\mathrm{OPT}_{r_1}$.
If $\boldsymbol{\zeta}^{(0)}$ is an optimal $r_0$-fold multicover, then
$\lceil r_1/r_0\rceil\boldsymbol{\zeta}^{(0)}$ is an $r_1$-fold multicover. Consequently,
$\mathrm{OPT}_{r_1}\leq\lceil r_1/r_0\rceil\mathrm{OPT}_{r_0}
\leq\lceil r_1/r_0\rceil\mathrm{OPT}^{\mathrm{risk}}_\eta$, which proves
the claimed token-cost bound.
\end{proof}

For fixed $p,\eta\in(0,1)$, we have $r_0=O(1)$ and $r_1=O(\log n)$, so
Theorem~\ref{thm:patch-policy-construction} gives an $O(\log^2 n)$ factor
relative to $\mathrm{OPT}^{\mathrm{risk}}_\eta$. If $p$ is fixed and
$1-\eta=n^{-\Theta(1)}$, then $r_0,r_1=\Theta(\log n)$ and
$\lceil r_1/r_0\rceil=O(1)$, improving the factor to $O(\log n)$.

A logarithmic dependence is unavoidable. Given an unweighted Set Cover
instance $(U,\{V_i\}_{i\in[k]})$, let oracle $i$ return $V_i$ with certainty,
unit token cost, and unit latency. With $p=\eta=1$, $y_e=1/n$, and $\rho=n$,
each response covering a new element satisfies local dual certification. A
one-turn batch is therefore successful exactly when its returned sets cover
$U$, and its token cost is the number of queried oracles. Thus, minimum-token
one-turn partial-cover selection inherits the Set Cover hardness: unless
$\mathrm P=\mathrm{NP}$, it has no polynomial-time
$(1-\varepsilon)\ln n$-approximation for any constant
$\varepsilon>0$~\cite{DinurSteurer2014}.

\section{Adaptive Selection with Partial-Cover Responses}
\label{sec:residual-cover}

We now consider multi-batch adaptive selection. Calls within a batch are
issued together, while the PTM may select each later batch using earlier
responses and may include the current residual $R$ in its queries. After
receiving a batch, the PTM may combine partial covers contained in several
responses and retains the resulting partial cover only if it satisfies the
local dual-certification condition. Response distributions may depend on $R$
and the preceding transcript; no independence among responses is required.

We derive a lower bound on the probability of returning a certified cover by
the deadline and an upper bound on expected token cost under a turn-by-turn
progress assumption.
Fix $\rho\geq1$, $\delta\in(0,1]$, and a turn bound $u$. For each $t\in[u]$,
fix $p_t\in(0,1]$, $c_t>0$, and $\ell_t>0$. Suppose the PTM adaptively
selects an oracle batch whenever turn $t$ is reached and $R\ne\varnothing$.
For every positive-probability history at such a turn, conditioning on that
history and the selected batch, assume the following: \textup{(1)} the
\emph{progress condition} holds with probability at least
$p_t$: the batch responses allow the PTM to retain a partial cover satisfying
the local dual-certification condition and covering at least $\delta|R|$
elements of $R$; \textup{(2)} the batch has conditional expected token cost
at most $c_t$; and \textup{(3)} the batch has latency at most $\ell_t$ almost
surely.

The parameters need not describe identical batches. One sufficient, but not
necessary, condition is that the joint response law of each selected batch
depends on the earlier transcript only through the current residual and,
for every reachable residual, yields such a partial cover with probability at
least $p_t$, expected token cost at most $c_t$, and latency at most $\ell_t$.
Thus, the parameters may be conservative envelopes over heterogeneous laws.

The PTM retains
such a partial cover whenever one is available, retains nothing from the batch
otherwise, and stops when the cover is complete. Let
$r=\min\{j\geq1:n(1-\delta)^j<1\}$; thus, $r$ turns satisfying the progress condition
suffice for completion. For comparison, let $B=\sum_{t=1}^u B_t$, where the
$B_t$ are independent Bernoulli random variables with means $p_t$.

\begin{lemma}
\label{lem:residual-progress}
If $\sum_{t=1}^u\ell_t\leq D$, then by $D$
the PTM returns a certified $\rho$-cover with probability at least
\(\Pr[B\geq r]\), and its expected token cost is at most
$\sum_{t=1}^u c_t$. \textup{(See
Appendix~\ref{app:proof-residual-progress} for a proof.)}
\end{lemma}

Lemma~\ref{lem:residual-progress} analyzes the specified adaptive policy but
does not yield a lower bound for other policies, because it places no upper
bound on the number of previously uncovered elements that another policy may
cover in one batch. For the upper bound below, a reached turn is
\emph{productive} if the PTM retains a nonempty partial cover satisfying the
local dual-certification condition; otherwise its retained-cover residual is
unchanged. For the lower bound, let $U_t^{\mathrm{seen}}\subseteq U$ be the
union of all valid candidate sets proposed in responses through turn $t$,
whether retained or not, with $U_0^{\mathrm{seen}}=\varnothing$.
A turn makes \emph{revealed progress} if
$U_t^{\mathrm{seen}}\setminus U_{t-1}^{\mathrm{seen}}\ne\varnothing$;
its revealed coverage is the size of this difference.
Thus, the lower bound permits buffering locally uncertified proposals and
certifying only their final union.

Fix integers $1\leq a\leq b\leq n$, probabilities
$\underline p,\overline p\in(0,1]$, token-cost bounds
$0<\underline c\leq\overline c$, and latency bounds
$0<\underline\ell\leq\overline\ell$. Put
$r_a=\lceil n/a\rceil$ and $r_b=\lceil n/b\rceil$. The
\emph{upper-policy conditions} hold if there is an adaptive selection-only
policy $\pi$ that retains only locally dual-certified partial covers and,
conditional on every reached positive-probability
history and each batch selected by $\pi$ with positive conditional
probability, the turn is productive with probability at least $\underline p$,
covers at least $\min\{a,|R|\}$ residual elements when productive, has
conditional expected token cost at most $\overline c$, and has latency at
most $\overline\ell$ almost surely. The \emph{universal batch limitations}
hold if, conditional on every reached positive-probability history and each
batch selected with positive conditional probability by any selection-only
PTM, the turn makes revealed progress with probability at most $\overline p$,
has revealed coverage at most $b$ when it makes progress, has conditional
expected token cost at least $\underline c$, and has latency at least
$\underline\ell$ almost surely. For
$u\overline\ell\leq D$, define the success-probability lower bound
$\eta_U(u)=\Pr[\operatorname{Bin}(u,\underline p)\geq r_a]$ and the
expected-token-cost upper bound
$\TOK_U(u)=\overline c\sum_{t=0}^{u-1}
\Pr[\operatorname{Bin}(t,\underline p)<r_a]
\leq\overline c\min\{u,r_a/\underline p\}$.
Here $\underline p,\overline p,\underline c,\overline c,
\underline\ell,\overline\ell$ are uniform envelopes over the relevant
histories and selected batches, not homogeneity assumptions on the HSOs.

\begin{theorem}
\label{thm:bounded-patch-sandwich}
\textup{(1)} Under the upper-policy conditions, for every integer $u$ with
$u\overline\ell\leq D$, running $\pi$ until it
completes or has used $u$ turns returns a certified $\rho$-cover by $D$ with
probability at least $\eta_U(u)$ and expected token cost at most $\TOK_U(u)$.

\textup{(2)} Under the universal batch limitations on revealed coverage, any
deadline- and turn-feasible selection-only PTM with turn bound
$u$ that returns a certified $\rho$-cover by $D$ with probability at
least $\eta$ has expected token cost at least
$\eta r_b\underline c/\overline p$. For such a PTM, with
$v=\min\{u,\lfloor D/\underline\ell\rfloor\}$, necessarily
$\eta\leq\Pr[\operatorname{Bin}(v,\overline p)\geq r_b]$.
\textup{(See Appendix~\ref{app:proof-bounded-patch-sandwich} for a proof.)}
\end{theorem}

Consequently, if both the upper-policy conditions and the universal batch
limitations hold and $\eta\leq\eta_U(u)$,
the instance-level token complexity lies between
$\eta r_b\underline c/\overline p$ and $\TOK_U(u)$. When $a=b$,
$\underline p=\overline p=p$, and $\underline c=\overline c=c$, the
bounds are within a factor $1/\eta$.

For a fixed feasible WSC instance and selection policy, the probability that
returned partial covers can be combined into a full cover depends on how
the oracle responses are jointly distributed; under adaptive selection, this
information is represented by the relevant history-conditional response
distributions. We do not assume that the PTM knows these distributions.

The ``all but one'' construction for stochastic Set Cover already shows a
linear adaptivity gap~\cite{GoemansVondrak2006}; round-limited stochastic
covering has also been studied~\cite{GhugeGuptaNagarajan2024}. We adapt that
pattern to a WSC instance with LP-dual certification, token costs, and a
two-turn deadline.

\textup{(1)} Fix $z\geq3$ and take $z$
disjoint promised yes-instances $(U_j,\mathcal A_j)$ of Exact Cover by 3-Sets,
where $|U_j|=3r_j$~\cite{GareyJohnson1979}. Let
$U=\mathbin{\dot\bigcup}_{j=1}^zU_j$, so
$n=|U|=3\sum_{j=1}^zr_j$. For every $A\in\mathcal A_j$, include $A$ as a WSC
candidate set of weight $1/(zr_j)$, and include $U_j$ as an additional
candidate set of weight $2/z$. No candidate set intersects two distinct
subuniverses. The candidates $U_j$ make the incidence graph have exactly $z$ connected
components, so the PTM can recover $(U_1,\ldots,U_z)$ from the WSC instance.
Index the candidate sets as $S_1,\ldots,S_m$, and set
$y_e=1/(3zr_j)$ for $e\in U_j$.

\textup{(2)} For each $j\in[z]$, fix $P_j\subseteq[m]$ indexing an exact cover
contained in $\mathcal A_j$; it contains $r_j$ candidate sets and has weight
$1/z$. The PTM is not given $P_j$. A specialist call opportunity indexed by
$j$ accepts only the WSC subinstance on $U_j$ and returns $P_j$, which the
PTM can verify. Put
$B_j=\bigcup_{\ell\ne j}P_\ell$. Neither $B_j$ nor $P_j$ covers $U$, whereas
$B_j\cup P_j$ does. A near-cover oracle returns $B_J$, where $J$ is a random
variable satisfying $\Pr[J=j]=p_j>0$ for every $j\in[z]$ and
$\sum_{j=1}^zp_j=1$. The available opportunities are exactly one near-cover
call and the $z$ specialist calls, each usable at most once. The PTM knows the
oracle roles, their guarantees, and the probabilities $p_j$, but not the
candidate sets that their responses will contain. The near-cover call costs
$c_0>0$, specialist call $j$ costs $c_j>0$, and every call has unit latency.

\textup{(3)} In the adaptive strategy, the PTM first queries the near-cover oracle. After
receiving $B_j$, it computes
$R=U\setminus\bigcup_{i\in B_j}S_i=U_j$ and asks the specialist to solve this
residual WSC subinstance. The one-turn alternative issues all $z$ specialist
queries in one batch. Set $D=2$. The following theorem gives the exact
one- and two-turn costs.

\begin{theorem}
\label{thm:residual-separation}
In this setting, $\mathbf y$ is dual feasible and has objective value
$\sum_{e\in U}y_e=1$, while $\bigcup_{j=1}^zP_j$ is an integral cover of
weight $1$; consequently, $L=\mathrm{OPT}=1$. Among
selection-only PTMs that return an optimal cover with probability $1$ by the
deadline, the minimum expected token cost is $\sum_{j=1}^z c_j$ with one turn and
$\min\{\sum_{j=1}^z c_j,\ c_0+\sum_{j=1}^z p_jc_j\}$ with two turns. If
$c_j=c(n)>0$ for every $j\in[z]$ and $c_0=\alpha c(n)$, where $c(n)$ is any
positive function and $\alpha>1$ is a constant, the one-turn/two-turn ratio is
$z/\min\{z,\alpha+1\}$. In particular, when $z>\alpha+1$, the ratio is
$z/(\alpha+1)$.
\end{theorem}
\begin{proof}
For a three-element candidate set $A\in\mathcal A_j$,
$\sum_{e\in A}y_e=1/(zr_j)=w(A)$, while
$\sum_{e\in U_j}y_e=1/z\leq2/z=w(U_j)$. Thus $\mathbf y$ is dual feasible
and $\sum_{e\in U}y_e=\sum_{j=1}^z3r_j/(3zr_j)=1$, so $L\geq1$. The
subfamily $\bigcup_{j=1}^zP_j$ covers $U$ with weight $1$, so
$\mathrm{OPT}\leq1$. Since $L\leq\mathrm{OPT}$, both equal $1$. Moreover,
$w(B_j)=(z-1)/z$ and
$w(P_j)=1/z$, so $B_j$ and then $P_j$ satisfy the local dual-certification
condition with $\rho=1$. Thus the PTM can complete an optimal cover either in
one turn by issuing all specialist queries in one batch or in two turns by
first obtaining $B_J$ and then obtaining $P_J$ from the specialist.

Condition first on a realization of the PTM's random source. In one turn,
if specialist call $j$ is omitted and the near-cover call is not issued, then
$U_j$ is uncovered. If the near-cover call is issued, then on the event
$J=j$, which has positive probability, its response $B_j$ also leaves $U_j$
uncovered. Probability-$1$ completion therefore requires all specialist
calls, and their total cost is $\sum_jc_j$.

With two turns, issuing the near-cover call first and then specialist call
$J$ costs $c_0+\sum_jp_jc_j$ in expectation; issuing all specialist calls in
one turn remains available. Conversely, suppose the first batch contains the
near-cover call and the specialist calls indexed by $A\subseteq[z]$. On the
event $J=j\notin A$, probability-$1$ completion requires specialist call $j$
in the second batch. The expected cost is therefore at least
$c_0+\sum_{j\in A}c_j+\sum_{j\notin A}p_jc_j
\geq c_0+\sum_jp_jc_j$. If the near-cover call is not issued in the first
batch, its response cannot determine another call in the same second batch;
thus probability-$1$ completion requires all specialist calls across the two
turns and costs at least $\sum_jc_j$. Since the PTM's overall success
probability is $1$, its conditional success probability is $1$ for almost
every realization of its random source. Applying the preceding deterministic
lower bounds and averaging shows that randomization cannot improve the smaller
of these two costs. For
$c_j=c(n)$ and $c_0=\alpha c(n)$, this gives costs $zc(n)$ and
$\min\{zc(n),(\alpha+1)c(n)\}$.
\end{proof}

This construction captures how an initial response can identify a residual
subproblem, allowing one targeted follow-up to replace nonadaptive queries for
all possible residuals.

\section{Conclusion}
\label{sec:conclusion}

We studied how a PTM can use heterogeneous stochastic oracles to compute
certified solutions for WSC by a prescribed completion deadline, and analyzed
the resulting expected token cost. An LP lower bound enables the PTM to verify instance-specific
approximation guarantees without trusting oracle claims. For full-cover
responses, we characterized the cost--success frontier, developed an FPTAS
for two-turn selection, and established the complexity of exact optimization.
For partial-cover responses, we obtained one-turn selection guarantees and
adaptive bounds, and exhibited an explicit family in which residual-guided
follow-up queries yield a linear token-cost advantage over nonadaptive
selection. Together with the pilot study, these results show how
verifiable oracle responses and adaptive selection can improve solutions on
particular instances despite the classical worst-case approximation barrier.
Learning unknown response distributions, token costs, and latencies during
selection is a natural direction for further study. An important practical
open problem is to evaluate the certified oracle-selection approach on
unreduced real-world instances, particularly the large OR-Library railway
instances. The PTM may retain the full instance locally and query oracles with
smaller residual, retrieved, or overlapping subinstances. Simply partitioning
the instance into disjoint subinstances can omit cross-boundary candidate sets
or duplicate costs, so any decomposition must preserve global feasibility and
approximation certificates.

\appendix
\section{Proofs Omitted from the Main Body}

\subsection{Proof of Proposition~\ref{prop:certified-set-cover}}
\label{app:proof-certified-set-cover}
\begin{proof}
Because $0<p_i<1$ and call successes are independent, every all-failure
prefix has positive probability, and a deterministic schedule
attains $p^*$ if and only if its all-failure path uses every call opportunity.
Thus, with one turn, the PTM must query $[k]$, incurring expected token cost
$c_{[k]}$. With at most two turns, a deterministic schedule is specified by a
nonempty first batch $A$, followed after failure by $[k]\setminus A$ when the
complement is nonempty; it meets
the deadline exactly when $\ell_A+\ell_{[k]\setminus A}\leq D$ and has
expected token cost $c_A+f_A\cdot c_{[k]\setminus A}$. A randomized policy
induces a distribution over these deterministic schedules. To attain $p^*$,
it can assign positive probability only to schedules attaining $p^*$, so its
expected token cost, being a weighted average of their costs, cannot be below
their minimum.
\end{proof}

\subsection{Proof of Theorem~\ref{thm:two-turn-exact-hardness}}
\label{app:proof-two-turn-hardness}
\begin{proof}
\textup{(1)} Reduce from \textsc{Partition}~\cite{GareyJohnson1979}. Let positive
integers $a_1,\ldots,a_k$ sum to $2B$, and put $M=4B^3$. Give oracle $i$
cost $c_i=a_i$, success probability $p_i=a_i/M$, and cutoff $\ell_i=1$;
these are valid probabilities because $a_i\leq2B$ and hence
$0<p_i\leq1/(2B^2)\leq1/2<1$.
Set $D=2$ and $\tau=2B-(B^2-1/2)/M$. Consider the decision version of
\textsc{2T-MaxSuccess}, which asks whether $c_2^*\leq\tau$. We show that its
answer is yes exactly when the \textsc{Partition} instance is a yes-instance.
Every nonempty subset is a feasible first batch, and maximizing the
expected saving is equivalent to minimizing the expected two-turn token cost
$c_A+f_A\cdot c_{[k]\setminus A}$.
For $s=\sum_{i\in A}a_i$, the union bound gives
$1-f_A\leq\sum_{i\in A}p_i=s/M$, while
$c_{[k]\setminus A}=2B-s$. Hence
\(F(A)=(1-f_A)c_{[k]\setminus A}\leq s(2B-s)/M\). If no subset sums to $B$,
then the integer $s$ differs from $B$ by at least $1$. Therefore,
$s(2B-s)=B^2-(s-B)^2\leq B^2-1$, and
$F(A)\leq(B^2-1)/M$ for every $A$.
If some subset $A$ sums to $B$, then
$x=\sum_{i\in A}p_i=M^{-1}\sum_{i\in A}a_i=B/M$ and
$c_{[k]\setminus A}=\sum_i a_i-\sum_{i\in A}a_i=2B-B=B$. The inequalities
$1-p_i\leq e^{-p_i}$ imply
$\prod_{i\in A}(1-p_i)\leq e^{-\sum_{i\in A}p_i}=e^{-x}$, which together with
$1-e^{-x}\geq x-x^2/2$ gives
$1-f_A\geq B/M-B^2/(2M^2)$. Since the second batch $[k]\setminus A$ has
expected token cost $B$, multiplying this bound by $B$ yields
\(F(A)\geq B^2/M-B^3/(2M^2)=(B^2-1/8)/M\).
Since $c_{[k]}=2B$, Proposition~\ref{prop:certified-set-cover} gives
$c_2^*=2B-\max_A F(A)$. In a no-instance,
$\max_A F(A)\leq(B^2-1)/M<(B^2-1/2)/M$, so $c_2^*>\tau$. In a
yes-instance, $\max_A F(A)\geq(B^2-1/8)/M>(B^2-1/2)/M$, so
$c_2^*<\tau$. Thus $c_2^*\leq\tau$ exactly when the \textsc{Partition}
instance is a yes-instance. Moreover, $\ell_i=1$ and
$p_i/c_i=(a_i/M)/a_i=1/M$ for every $i$. The constructed rational parameters
have encoding length polynomial in the \textsc{Partition} input length, so
the reduction runs in polynomial time. This proves that the decision version
is NP-hard, and so \textsc{2T-MaxSuccess} is NP-hard.

\textup{(2)} Enumerate $a\in\{\ell_1,\ldots,\ell_k\}$ and
$b\in\{0,\ell_1,\ldots,\ell_k\}$ with $a+b\leq D$.
Set $L_{a,b}=\{i:\ell_i>b\}$ and $U_a=\{i:\ell_i\leq a\}$, discarding pairs
for which $L_{a,b}\nsubseteq U_a$. Each remaining pair defines the family
$\mathcal A_{a,b}=\{A\subseteq[k]:L_{a,b}\subseteq A\subseteq U_a\}$.
Set $L=L_{a,b}$ and $S=U_a\setminus L$.
Every $A\in\mathcal A_{a,b}$ has the form $A=L\cup X$ for a unique
$X\subseteq S$; the state $L\cup X=\varnothing$ is discarded. A subset-sum
dynamic program processes the elements $i\in S$. Each state
records a cost $s$ and the smallest product $\prod_{i\in X}(1-p_i)$ attained
by a processed subset $X$ of that cost: excluding $i$ preserves both values,
whereas including it adds $c_i$ to the cost and multiplies the product by
$1-p_i$. Backpointers retain a corresponding subset. Thus, for every integer
$s\in[0,c_{[k]}]$, the program stores the smallest product among subsets
$X\subseteq S$ such that $c_X=s$. For fixed $s$, this
product minimizes $f_{L\cup X}$, while the complementary cost
$c_{[k]}-c_L-s$ is fixed; hence it maximizes
\(F(L\cup X)=(1-f_{L\cup X})(c_{[k]}-c_L-s)\).
Taking the best stored state over all costs and all timeout pairs gives
the exact optimum. There are $O(k^2)$ pairs and $O(kc_{[k]})$ states per pair;
the rational products have polynomial bit length. Thus the algorithm is
pseudopolynomial in $c_{[k]}$ and runs in polynomial time when the
integer costs are polynomially bounded.
\end{proof}

\subsection{Proof of Lemma~\ref{lem:patch-composition}}
\label{app:proof-patch-composition}
\begin{proof}
The sets $N(P_h,R_{h-1})$ of newly covered elements are pairwise disjoint,
and their union is $U$ when the cover is complete. Since
$w(P_h)\leq\rho\sum_{e\in N(P_h,R_{h-1})}y_e$ for every $h$ and the set
weights are nonnegative, $w(\mathcal C)\leq\sum_{h=1}^H w(P_h)
\leq\rho\sum_{e\in U}y_e\leq\rho L\leq\rho\,\mathrm{OPT}$.
\end{proof}

\subsection{Proof of Lemma~\ref{lem:residual-progress}}
\label{app:proof-residual-progress}
\begin{proof}
Every turn satisfying the progress condition shrinks the number of uncovered
elements by a factor of at most $1-\delta$; other turns leave the retained
cover unchanged. After $r$ such turns, the residual has cardinality at most
$n(1-\delta)^r<1$ and hence is empty.

Let $\mathcal F_t$ denote all information available after the PTM selects the
batch for turn $t$ but before receiving its responses. On every reached turn,
let $q_t$ be the conditional probability, given $\mathcal F_t$, that the
progress condition holds; by assumption, $q_t\geq p_t$ almost surely. Extend
the process after completion by declaring every remaining turn successful and
setting $q_t=1$. Construct an equivalent process recursively using independent
random variables $U_t$ uniform on $[0,1]$: conditional on $\mathcal F_t$, its
progress indicator is $X_t=\mathbf 1\{U_t\leq q_t\}$, with the remaining
response information sampled according to its conditional distribution. Set
$B_t=\mathbf 1\{U_t\leq p_t\}$. Then $B_t\leq X_t$ almost surely, while the
$B_t$ are independent Bernoulli random variables with means $p_t$. If at least
$r$ of the $X_t$ equal $1$, either the process completed before some of those
indicators were declared successful or $r$ reached turns satisfied the
progress condition; in either case, the process completed. Therefore,
$\Pr[\text{completion by turn }u]\geq\Pr[B\geq r]$.

Each retained proposal satisfies the local dual-certification condition, so
Lemma~\ref{lem:patch-composition} certifies the returned cover. At most $u$
batches are issued, and their latency bounds imply completion or failure by
$\sum_t\ell_t\leq D$. For the token-cost bound, let $I_t$ indicate that turn
$t$ is reached, and let $\TOK_t$ be the token cost of its selected batch. By
the tower property and the conditional cost assumption,
$\E[I_t\TOK_t]=\E[I_t\E[\TOK_t\mid\mathcal F_t]]
\leq c_t\Pr[I_t=1]$. Hence
$\E[\TOK_M]=\sum_{t=1}^u\E[I_t\TOK_t]
\leq\sum_{t=1}^u c_t\Pr[I_t=1]\leq\sum_{t=1}^u c_t$.
\end{proof}

When $p_t=p$ for all $t$, the comparison variable $B$ has distribution
$\operatorname{Bin}(u,p)$.

\subsection{Proof of Theorem~\ref{thm:bounded-patch-sandwich}}
\label{app:proof-bounded-patch-sandwich}
\begin{proof}
\textup{(1)} Let $X_t$ indicate that reached turn $t$ is productive,
and extend the process after completion by setting $X_t=1$. Let
$\mathcal F_t$ contain the information available after $\pi$ selects its
batch for turn $t$ but before receiving the responses, and let
$q_t=\Pr[X_t=1\mid\mathcal F_t]$. The upper-policy conditions give
$q_t\geq\underline p$ almost surely. As in the uniform-variable coupling in
Appendix~\ref{app:proof-residual-progress}, construct the process recursively
using independent random variables $U_t$ uniform on $[0,1]$, with
$X_t=\mathbf 1\{U_t\leq q_t\}$, and put
$Z_t=\mathbf 1\{U_t\leq\underline p\}$. Then $Z_t\leq X_t$ almost surely,
and the $Z_t$ are independent Bernoulli random variables with mean
$\underline p$. Each productive turn covers at least $a$ new elements unless
fewer than $a$ remain, so $r_a$ productive turns complete the cover. Hence
$\Pr[\text{completion by turn }u]\geq
\Pr[\sum_{t=1}^uZ_t\geq r_a]=\eta_U(u)$, and
Lemma~\ref{lem:patch-composition} certifies the completed cover.

Turn $t+1$ is reached only if fewer than $r_a$ productive turns have occurred
among the first $t$ turns. The same coupling therefore bounds its reach
probability by $\Pr[\operatorname{Bin}(t,\underline p)<r_a]$. Multiplying
these reach probabilities by the conditional cost bound and summing gives
$\TOK_U(u)$. The probability sum in $\TOK_U(u)$ is the expected value of a
negative-binomial stopping time truncated at $u$ and is at most both $u$ and
the untruncated mean $r_a/\underline p$. The latency bound gives total latency
at most $u\overline\ell\leq D$.

\textup{(2)} Let $I_t$ indicate that turn $t$ is reached, let $Y_t$
indicate that it is reached and makes revealed progress, and let
$T=\sum_{t=1}^u I_t$ be the number of turns issued. A selection-only PTM
can return a full cover only if the sets proposed by its calls cover $U$.
Since one turn reveals coverage of at most $b$ elements, completion requires
at least $r_b$ turns making revealed progress. Hence
$\E[\sum_{t=1}^uY_t]\geq\eta r_b$. Let $\mathcal F_t$ contain the
information available after the batch for turn $t$ is selected but before its
responses are received. The universal batch limitations and the tower
property give
$\E[Y_t]=\E[I_t\Pr[Y_t=1\mid\mathcal F_t]]
\leq\overline p\,\E[I_t]$. Summing over $t$ yields
$\E[T]\geq\eta r_b/\overline p$.

Let $\TOK_t$ be the token cost of the batch at turn $t$ when that turn is
reached. Another application of the tower property gives
$\E[I_t\TOK_t]=\E[I_t\E[\TOK_t\mid\mathcal F_t]]
\geq\underline c\,\E[I_t]$. Therefore,
$\E[\TOK_M]=\sum_{t=1}^u\E[I_t\TOK_t]
\geq\underline c\,\E[T]
\geq\eta r_b\underline c/\overline p$.

By the latency lower bound, no more than
$v=\min\{u,\lfloor D/\underline\ell\rfloor\}$ turns can finish by $D$.
Set $Y_t=0$ for $t>v$ and after the PTM stops. Let
$q_t=\Pr[Y_t=1\mid\mathcal F_t]$, taking $q_t=0$ when turn $t$ is not
reached. Then $q_t\leq\overline p$ almost surely. Using independent uniform
variables $U_t$, construct the process recursively with
$Y_t=\mathbf 1\{U_t\leq q_t\}$ and define
$Z_t=\mathbf 1\{U_t\leq\overline p\}$. Thus, $Y_t\leq Z_t$ almost surely,
and the $Z_t$ are independent Bernoulli random variables with mean
$\overline p$. Consequently, $\sum_{t=1}^vY_t$ is dominated by
$\operatorname{Bin}(v,\overline p)$. Since completion requires at least
$r_b$ turns making revealed progress, the binomial tail gives the claimed necessary
probability bound.
\end{proof}

\section*{AI Methodology}
The author used OpenAI Codex to assist with brainstorming, identifying
potential flaws in author-developed mathematical arguments, improving
exposition, and editing LaTeX. This assistance materially affected the
presentation throughout the manuscript. Anthropic Claude Code was used as an
experimental AI system in the pilot study described in
Section~\ref{sec:empirical}. The research questions and central conceptual
contributions originated with the author. The author verified all content,
including the mathematical results, proofs, citations, references, and
experimental findings.

\clearpage
\bibliography{reference}

\end{document}